\documentclass[preprint,showpacs,preprintnumbers,amsmath,amssymb]{revtex4}

\usepackage{graphicx}
\usepackage{dcolumn}
\usepackage{bm}

\begin{document}

\preprint{}

\title{The sextic oscillator as a $\gamma$-independent potential}

\author{G. L\'evai}%
\email{levai@atomki.hu}

\affiliation{%
Institute of Nuclear Research of the Hungarian Academy of 
Sciences (ATOMKI), P. O. Box 51, H-4001 Debrecen, Hungary
}%

\author{J. M. Arias}
\email{pepe@nucle.us.es}
\affiliation{Departamento de F\'{\i}sica At\'omica, Molecular y
  Nuclear. Facultad de F\'{\i}sica, Universidad de Sevilla, Apartado 1065, 
41080 Sevilla, Spain  
}%

\begin{abstract}
The sextic oscillator is proposed as a two-parameter solvable
$\gamma$-independent potential 
in the Bohr Hamiltonian. It is shown that closed analytical expressions can be 
derived for the energies and wavefunctions of the first few levels and 
for the strength of electric quadrupole transitions between them. 
Depending on the parameters this potential has a minimum at $\beta=0$ 
or at $\beta>0$, and might also have a local maximum before reaching its 
minimum. A comparison with the spectral properties of the infinite square 
well and the $\beta^4$ potential is presented, together with a brief 
analysis of the experimental spectrum and E2 transitions of the 
$^{134}$Ba nucleus. 
\end{abstract}



\maketitle

\section{Introduction}

In the last few years there has been considerable interest in looking
for analytic solutions of the Bohr Hamiltonian which describes the collective
motion in nuclei  
in terms of shape variables ($\beta$, $\gamma$) \cite{bm}. This 
was initiated with the introduction of the interacting boson model (IBM)
\cite{ibm}. This model present four different dynamical symmetries each
one associated to a well defined nuclear shape. The model Hamiltonian
provides a natural way of going from a phase to another one by
changing systematically few parameters and, consequently
allows to study shape phase transitions. Thus, the IBM phase diagram
has been analyzed from different points of view
\cite{Die,Feng,Cas,Jo1,Ar3}. Specially important are the
critical points since in these situations structural changes occur
rapidly and is difficult design appropriate theoretical models. 
Recently, Iachello has proposed a new dynamical symmetry for
describing the critical point at the transition from spherical to
deformed $\gamma$-unstable shapes \cite{fi00}. This symmetry has been
called E(5) and is expected to occur 
in nuclei in which the $V(\beta,\gamma)$ potential depends only on the 
$\beta$ variable, and it has a relatively flat shape in the $\beta$ 
variable. Around this critical point of the phase transition the 
potential appearing in the Bohr Hamiltonian can be approximated by
an infinite square well in the $\beta$ variable.
In this case the Bohr Hamiltonian can be solved exactly in terms of 
Bessel functions, and various quantitative predictions can be obtained 
for various spectroscopic properties (ratios of the excitation energies 
and $B({\rm E}2)$ transitions), on the basis of which one can search for 
candidates for the E(5) symmetry among nuclei. 
Lately, other two dynamical symmetries X(5) \cite{fi01} and Y(5)
\cite{fi03} to describe the critical point at the transition from
spherical to axially deformed shapes and from axially deformed to
triaxial shapes, respectively, have been proposed. 

The introduction of the E(5) symmetry renewed the interest in studying exactly 
solvable $V(\beta,\gamma)$ potentials. Most efforts have concentrated on 
$\gamma$-independent potentials, for which the most well-known example 
is the harmonic oscillator in the five-dimensional space \cite{wj56} and 
its extension which contains a term proportional to $\beta^{-2}$ 
\cite{elliott}. More recently two more exactly solvable $\gamma$-unstable 
potentials have been discussed: the Coulomb and the Kratzer potentials 
\cite{fortunato}. Here again the latter potential is an extension of the 
former in the sense that it contains a term proportional to $\beta^{-2}$, 
which formally changes the $\tau$ variable, the analogue of the $l$ 
angular momentum of radial potentials in three spatial dimensions. 
This formal changing of $\tau$ introduces a minimum of the potentials 
at $\beta>0$ in both cases \cite{elliott,fortunato}. The bound solutions 
of the harmonic oscillator and the Coulomb potentials (and their extensions) 
are given in terms of generalized Laguerre polynomials \cite{as70}. 

Similarly to the three-dimensional case, these examples, together with 
the infinite square well, practically exhaust those exactly solvable 
potentials that contain a $\beta^{-2}$ term. Here we propose another 
potential which has this property, and although it is not exactly solvable 
in the classical sense, it has a number of features that make it an 
ideal potential to be used in the Bohr Hamiltonian. This is the sextic 
oscillator, which belongs to the class of quasi-exactly solvable (QES) 
potentials \cite{qes}. These potentials have the property that their 
solutions can be obtained in closed form 
for a number of energy eigenvalues, i.e. for the 
lowest few values of the $n$ principal quantum number, which are also 
the lowest in energy. This is clearly sufficient for potentials appearing 
in the Bohr Hamiltonian: it is rarely necessary to consider more than 
a few levels with the same angular momentum, and these can be obtained 
from the lowest few solutions of the sextic oscillator. Furthermore, 
the sextic oscillator has a more flexible shape than other solvable 
potentials, as depending on its parameters, it can have a minimum 
at $\beta=0$ or at $\beta=\beta_{\rm min}>0$, and in addition, it can 
also have a local maximum at $\beta_{\rm max}<\beta_{\rm min}$. 

The paper is structured as follows. In Sect. 2 the Bohr Hamiltonian is
revised together with its solutions for $\gamma$-independent
potentials. In Sect. 3 the lowest energy solutions of the Bohr
Hamiltonian for the sextic oscillator potential are worked
out. Section 4 is devoted to show the simple use of the sextic
oscillator as a flexible $\gamma$-independent potential. Finally, in
Sect. 5 we present preliminary applications and discuss future extensions.

\section{The Bohr Hamiltonian for $\gamma$-independent potentials}

Let us first consider the Bohr Hamiltonian describing the collective 
motion of a deformed nucleus 
in the five-dimensional space determined by the $\theta_i$ Euler angles 
($i=1,\ 2,\ 3$) and the intrinsic $\beta$ and $\gamma$ variables is 
\cite{bm}
\begin{equation}
H=-\frac{\hbar^2}{2B}\left(
\frac{1}{\beta^4}\frac{\partial}{\partial \beta}
\beta^4\frac{\partial}{\partial \beta}
+\frac{1}{\beta^2 \sin 3\gamma}\frac{\partial}{\partial \gamma}
\sin 3\gamma\frac{\partial}{\partial \gamma}
-\frac{1}{4\beta^2}\sum_k \frac{Q^2_k}{\sin^2(\gamma-\frac{2}{3}\pi k)}
\right)+V(\beta,\gamma)\ .
\label{bohr}
\end{equation}
In what follows we assume that the potential in (\ref{bohr}) depends 
only on $\beta$, i.e. $V(\beta,\gamma)=U(\beta)$. For these
$\gamma$-independent potentials the wavefunctions can be separated into two 
parts  
\begin{equation}
\Psi(\beta,\gamma,\theta_i)=f(\beta)\Phi(\gamma,\theta_i)\ ,
\label{fbgt}
\end{equation}
which satisfy the following differential equations:
\begin{eqnarray}
\left(-\frac{1}{\sin 3\gamma}\frac{\partial}{\partial \gamma}
\sin 3\gamma\frac{\partial}{\partial \gamma}
+\frac{1}{4}\sum_k \frac{Q^2_k}{\sin^2(\gamma-\frac{2}{3}\pi k)}
\right)\Phi(\gamma,\theta_i)=\Lambda\Phi(\gamma,\theta_i)\ ,
\nonumber\\
\Lambda=\tau(\tau+3), \hskip .5cm \tau=0,\ 1,\ 2,\ \dots
\label{bohgt}
\end{eqnarray}
\begin{equation}
\left(
-\frac{1}{\beta^4}\frac{\partial}{\partial \beta}
\beta^4\frac{\partial}{\partial \beta}
+\frac{\Lambda}{\beta^2}+u(\beta)\right)f(\beta)=\epsilon f(\beta)\ .
\label{bohrb}
\end{equation}
Here we have introduced $\epsilon=\frac{2B}{\hbar^2}E$ and 
$u(\beta)=\frac{2B}{\hbar^2}U(\beta)$. Note that the $\tau$ values 
determine the allowed angular momenta $J$ too \cite{bes}.
By setting 
$\phi(\beta)=\beta^2 f(\beta)$ we obtain an equation which has the 
form of a radial Schr\"odinger equation
\begin{equation}
-\frac{{\rm d}^2\phi}{{\rm d}\beta^2}
+\left(\frac{(\tau+1)(\tau+2)}{\beta^2}+u(\beta)
\right)\phi=\epsilon\phi\ .
\label{bohrc}
\end{equation}
Note that this is different from the Eq. (6) in \cite{fi00} in that 
it contains no linear derivative term due to the different definition 
of $\phi(\beta)$. This choice also implies that the factor corresponding 
to the $\beta$ volume element in the integration of functions of the type 
$f(\beta)$ in (\ref{bohrb}) has been transferred to the solutions of 
the type $\phi(\beta)$ in (\ref{bohrc}). Thus, in the integration of 
these no factor arising from the volume element has to be included. 
The complete solution of the problem implies the solution of Eq. 
(\ref{bohgt}) too, this was solved in Ref. \cite{bes}. 

\section{The sextic oscillator}

The sextic oscillator with a centrifugal barrier is defined \cite{qes} 
as 
\begin{equation}
H=-\frac{{\rm d}^2 }{{\rm d} x^2}+\frac{(2s-1/2)(2s-3/2)}{x^2}
+\left(b^2-4a(s+\frac{1}{2}+M)\right)x^2+2abx^4+a^2x^6\ ,
\label{sexpot}
\end{equation}
where $x\in[0,\infty)$ and $M$ is a non-negative integer. This potential 
is quasi-exactly solvable, which means that for any non-negative integer 
value of $M$, $M+1$ of its solutions can be obtained in an algebraic 
way. The (unnormalized) solutions are written in the form 
\begin{equation}
\phi_n(x)=P_n(x^2)(x^2)^{s-\frac{1}{4}}
\exp\left(-\frac{a}{4}x^4-\frac{b}{2}x^2\right)\,    
\hskip .5cm n=0,\ 1,\ 2,\ \dots
\label{sexpol}
\end{equation}
where $P_n$ is a polynomial of order $n$. Obviously, normalizability 
requires $a\ge 0$, while $a=0$ reduces the problem to the exactly solvable 
harmonic oscillator. 

The simplest solutions are obtained for $M=0$ and $M=1$
\cite{qes}. For $M=0$ only one  
nodeless (i.e. ground-state) solution appears at $E^{(M=0)}_0=4bs$, with 
the corresponding wavefunction being
\begin{equation}
\phi^{(M=0)}_0(x)\sim (x^2)^{s-\frac{1}{4}}
\exp\left(-\frac{a}{4}x^4-\frac{b}{2}x^2\right)\ . 
\label{sexwf0}
\end{equation}
For $M=1$ two solutions appear, one nodeless, and another with one 
node for $x>0$. These correspond to the ground-state and the first 
excited state, respectively, at energies $E^{(M=1)}_0=4bs+\lambda_-(s)$ and 
$E^{(M=1)}_1=4bs+\lambda_+(s)$, where 
\begin{equation}
\lambda_{\pm}(s)=2b\pm 2(b^2+8as)^{1/2}
\label{lams}
\end{equation}
are the roots of the equation $\lambda^2-4b\lambda-32as=0$. The 
corresponding wavefunctions are
\begin{equation}
\phi^{(M=1)}_n(x)\sim \left(1-\frac{\lambda}{8s}x^2\right)
(x^2)^{s-\frac{1}{4}}\exp\left(-\frac{a}{4}x^4-\frac{b}{2}x^2\right)\ ,
\label{sexwf1}
\end{equation}
and the $\lambda=\lambda_-(s)$ and $\lambda=\lambda_+(s)$ choice has
to be made 
for $n=0$ and $n=1$, respectively \cite{qes}. (Note that $a\ge 0$ and 
$s\ge 0$  imply $\lambda_-(s)\le 0$, so the polynomial part
of (\ref{sexwf1}) is  
nodeless.) It has to be mentioned that the solutions for $M=0$ and $M=1$ 
belong to {\it different} sextic potentials if $s$ is the same, as the 
coefficient of the quadratic term is different then. We shall see, 
however, that with appropriate combinations of $s$ and $M$ it is possible 
to solve sextic potentials that differ only in the strength of the 
centrifugal term. 

The normalization of the wavefunctions can also be given in closed form. 
For this one has to evaluate integrals of the type 
\begin{eqnarray}
I^{(A)}&\equiv&\int_0^{\infty} x^A \exp\left(-\frac{a}{2}x^4-b x^2\right)
\nonumber\\
&=&\frac{1}{2}\Gamma\left(\frac{A+1}{2}\right)a^{-\frac{A+1}{4}}\exp\left(\frac{b^2}{4a}\right)
D_{-\frac{A+1}{2}}\left(\frac{b}{a^{1/2}}\right)
\label{sexintd}\\
&=&\frac{1}{2}\Gamma\left(\frac{A+1}{2}\right)(2a)^{-\frac{A+1}{4}}
U\left(\frac{A+1}{4},\frac{1}{2};\frac{b^2}{2a}\right)
\label{sexintu1}\ ,
\end{eqnarray}
where $D_p(z)$ is the parabolic cylinder function and $U(\alpha,\beta;z)$ 
is one of the forms of the confluent hypergeometric function 
\cite{as70}. 

Larger values of $M$ can 
also be considered (e.g. for $M=2$ three different solutions are obtained 
for the three roots of a cubic algebraic equation for $\lambda$), but 
$M=0$ and $M=1$ are sufficient for our purposes in this paper. A
complete study including more solutions, explicit closed forms for the
normalization factors and applications to actual
nuclei is underway. 

\section{Application as a $\gamma$-independent potential}

In order to cast (\ref{sexpot}) in a form similar to (\ref{bohrc}) we 
have to write $x=\beta$ and $s=\frac{\tau}{2}+\frac{5}{4}$ (remember 
that $\tau\ge 0$). In order to keep the quadratic term at a constant value 
(once the $a,b$ parameters are fixed) we also have to prescribe 
\begin{equation}
s+M+\frac{1}{2}=\frac{1}{2}\left(\tau+2M+\frac{7}{2}\right)\equiv c=const.  
\label{abc}
\end{equation}
With this the sextic oscillator Hamiltonian can be brought to the form 
of (\ref{bohrb}) with $u(\beta)$ being 
\begin{equation}
u^{\pi}(\beta)=(b^2-4ac^{\pi})\beta^2+2ab\beta^4+a^2\beta^6+u_0^{\pi}\ ,
\label{upi}
\end{equation}
where the index $\pi=\pm$ is included to distinguish the potential for
even/odd $\tau$'s which is slightly different as explained below. In
Eq. (\ref{upi}) $c^{\pi}$ are the constants obtained in Eq. (\ref{abc})
for even/odd values of $\tau$ and we have introduced a constant
$u_0^{\pi}$ for convenience as it will be discussed below. 

Equation (\ref{abc}) implies that increasing/decreasing $M$ in one unit 
has to come with decreasing/increasing $\tau$ in {\it two}
units. Thus, once the values of the $(a,b)$ parameters are fixed, the
sequence of $(M,\tau)$ values 
$(K,0),(K-1,2),(K-2,4),\dots$ correspond to solutions of
Eq. (\ref{upi}) with  $c^{+}=\frac{7}{4}+K$. In the same way, the
sequence of $(M,\tau)$ values 
$(K,1),(K-1,3),(K-2,5),\dots$ correspond to solutions of
Eq. (\ref{upi}) with  $c^{-}=\frac{9}{4}+K$.
Consequently, the potential for $\tau$-even and 
$\tau$-odd states is slightly different due to the fact that the 
coupling coefficient $b^2-4ac^{\pm}$ of the quadratic term is different 
in the two cases due to the choices for $c^+$ and $c^-$ that are 
necessary to separate the $(\tau+1)(\tau+2)\beta^{-2}$ term in a uniform 
way. This situation can be handled using different strategies. One 
possibility is setting $u_0^+=u_0^-=0$ and considering $b^2 > 10 a$, 
which minimizes the deviation of the quadratic terms compared to the 
quartic and sextic terms. Another possibility is introducing a relative 
energy shift between the $\tau$-even and $\tau$-odd potentials 
by setting $u_0^+$ and $u_0^-$ such that the potential minima are
at the same energy. We shall discuss this 
possibility after analyzing qualitatively the spectrum and 
the potential shapes. 

\begin{figure}
\resizebox{10cm}{!}{\rotatebox{0}{\includegraphics{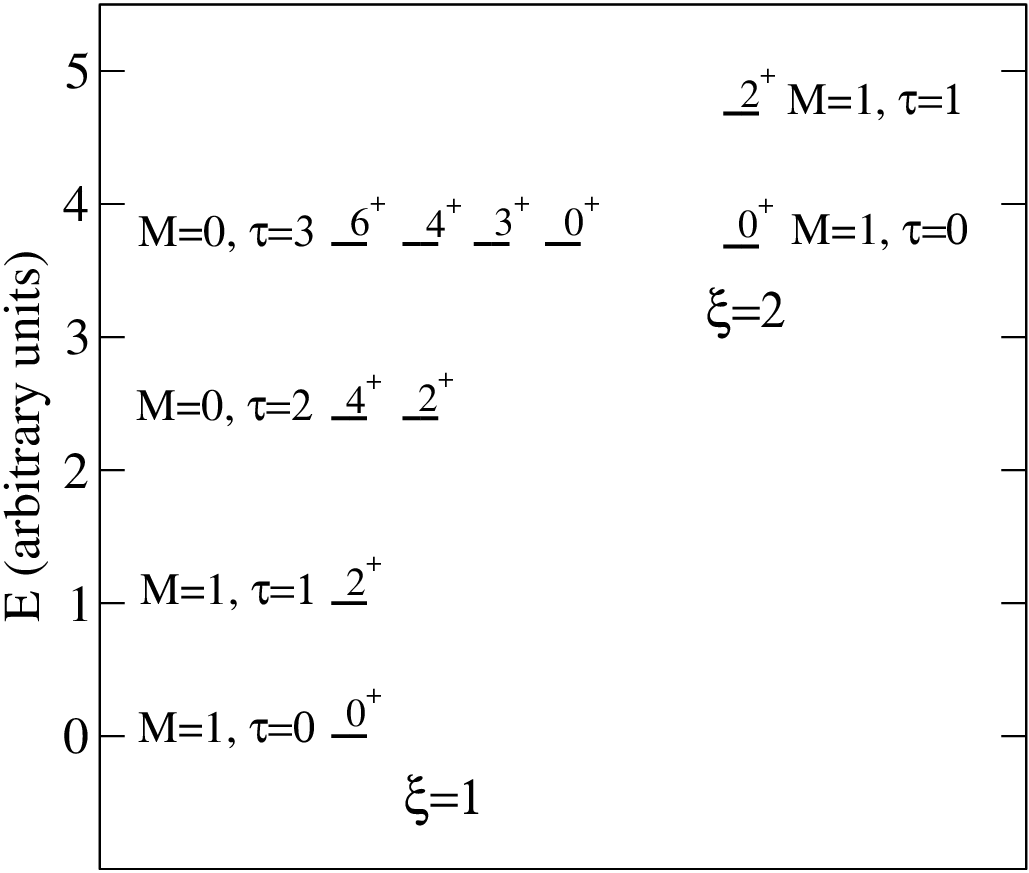}}}
\caption{Schematic typical spectrum for the sextic oscillator with
  indication of the relevant quantum numbers.}
\label{spectrum}
\end{figure}

\begin{figure}
\resizebox{10cm}{!}{\rotatebox{0}{\includegraphics{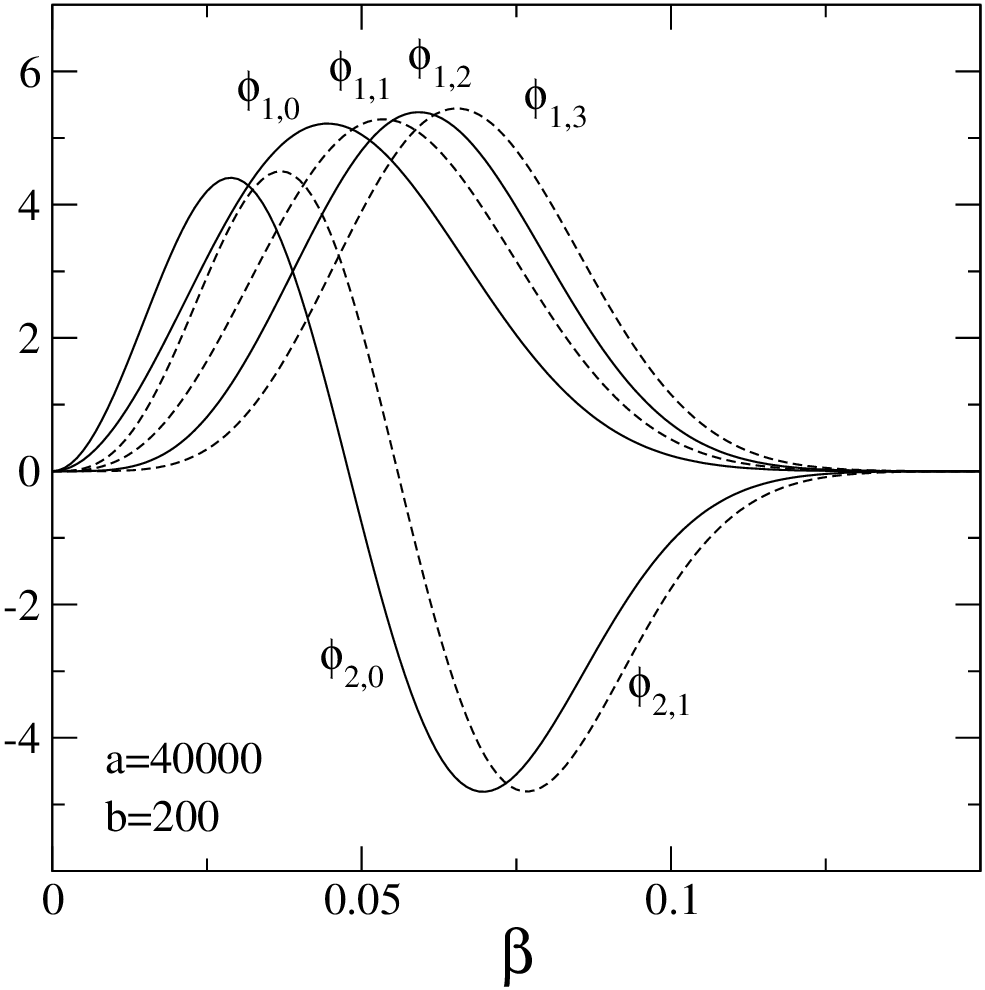}}}
\caption{Wavefunctions with the notation $\phi_{\xi,\tau}$ for the
  case of potential parameters $a=40000$ and $b=200$.}
\label{wfunctions}
\end{figure}

Let us analyze the spectrum obtained in a simple case. Taking $M=1$ we
obtain, as explained in the preceding section, two solutions for
$\tau=0$ (one with no nodes, $n=0$, and the other one with one node,
$n=1$). In the 
notation introduced in Ref. \cite{fi00} the label $\xi$ is our $n+1$.
Thus the two solutions of Eq. (\ref{sexwf1}) with $s=5/4$ are
$\phi^{(M=1)}_n(\beta)$ 
with $n=0$ and $1$ and correspond to $\phi_{\xi,\tau}=\phi_{1,0}$ and
$\phi_{\xi,\tau}=\phi_{2,0}$ respectively in Ref. \cite{fi00} notation. 
Inspecting  the energy eigenvalues, the corresponding energies are
$E_1(M=1,\tau=0)=E_{1,0}=7b - 2(b^2+10a)^{1/2}+u_0^+$ and 
$E_2(M=1,\tau=0)=E_{2,0}=7b + 2(b^2+10a)^{1/2}+u_0^+$, respectively. 
The same potential is obtained by taking $M=0$ and $\tau=2$. In this
case we have a single solution, Eq. (\ref{sexwf0}) with $s=9/4$, 
$\phi^{(M=0)}_0(\beta)$
with no nodes and with energy $E_1(M=0,\tau=2)=9b+u_0^+$. This
corresponds to $\phi_{\xi,\tau}=\phi_{1,2}$ in the notation of
Ref. \cite{fi00}.
A similar analysis can be performed for the solutions with odd
$\tau$-values. For $M=1$ there are two solutions with $\tau=1$,
Eq. (\ref{sexwf1}) with $s=7/4$, which correspond to
$\phi_{\xi,\tau}=\phi_{1,1}$ and  $\phi_{\xi,\tau}=\phi_{2,1}$  for
$n=0$ and $n=1$ respectively.  The corresponding energy 
eigenvalues  are $E_{1,1}=9b-2(b^2+14a)^{1/2}+u_0^-$ 
and $E_{2,1}=9b+2(b^2+14a)^{1/2}+u_0^-$. Again the same potential is
obtained for $M=0$ and $\tau=3$. In this case there is a single
solution with no nodes, Eq. (\ref{sexwf0}) with $s=11/4$,  
which corresponds to $\phi_{1,3}$ and has an energy
$E_{1,3}=11b+u_0^-$. In Fig. \ref{spectrum} a schematic spectrum is
shown with indication of the relevant quantum numbers. In
Fig. \ref{wfunctions} the corresponding wavefunctions with the notation 
$\phi_{\xi,\tau}$ are presented.

Now we analyze the different potential shapes that can be
produced by different election of parameters in Eq. (\ref{upi}). 
From (\ref{upi}) we find that the shape of the potential $u^{\pi}(\beta)$ 
depends on the sign of $b^2-4ac^{\pi}$ and $b$, which set the coefficients 
of the quadratic and quartic terms. (The coefficient of the leading sextic 
term is always positive.) When $b^2>4ac^{\pi}$ and $b>0$ hold (i.e. for 
$b>2(ac^{\pi})^{1/2}$), the potential 
has a minimum at $\beta=0$ and it increases monotonously with $\beta$. 
When $b^2<4ac^{\pi}$, irrespective of the sign of $b$ (i.e. for 
$-2(ac^{\pi})^{1/2} < b < 2(ac^{\pi})^{1/2}$), a minimum appears for 
$\beta>0$, while for $b^2>4ac^{\pi}$ and $b<0$ (i.e. for 
$b<-2(ac^{\pi})^{1/2}$), first a maximum appears 
and then a minimum as $\beta$ increases. In all three cases the exact 
location of the extremal point(s) can be obtained from the real and 
positive solutions of
\begin{equation}
(\beta^{\pi}_0)^2=\frac{1}{3a}[-2b\pm(b^2+12ac^{\pi})^{1/2}]\ .
\label{beta0}
\end{equation}

\begin{figure}
\resizebox{\columnwidth}{!}{\rotatebox{0}{\includegraphics{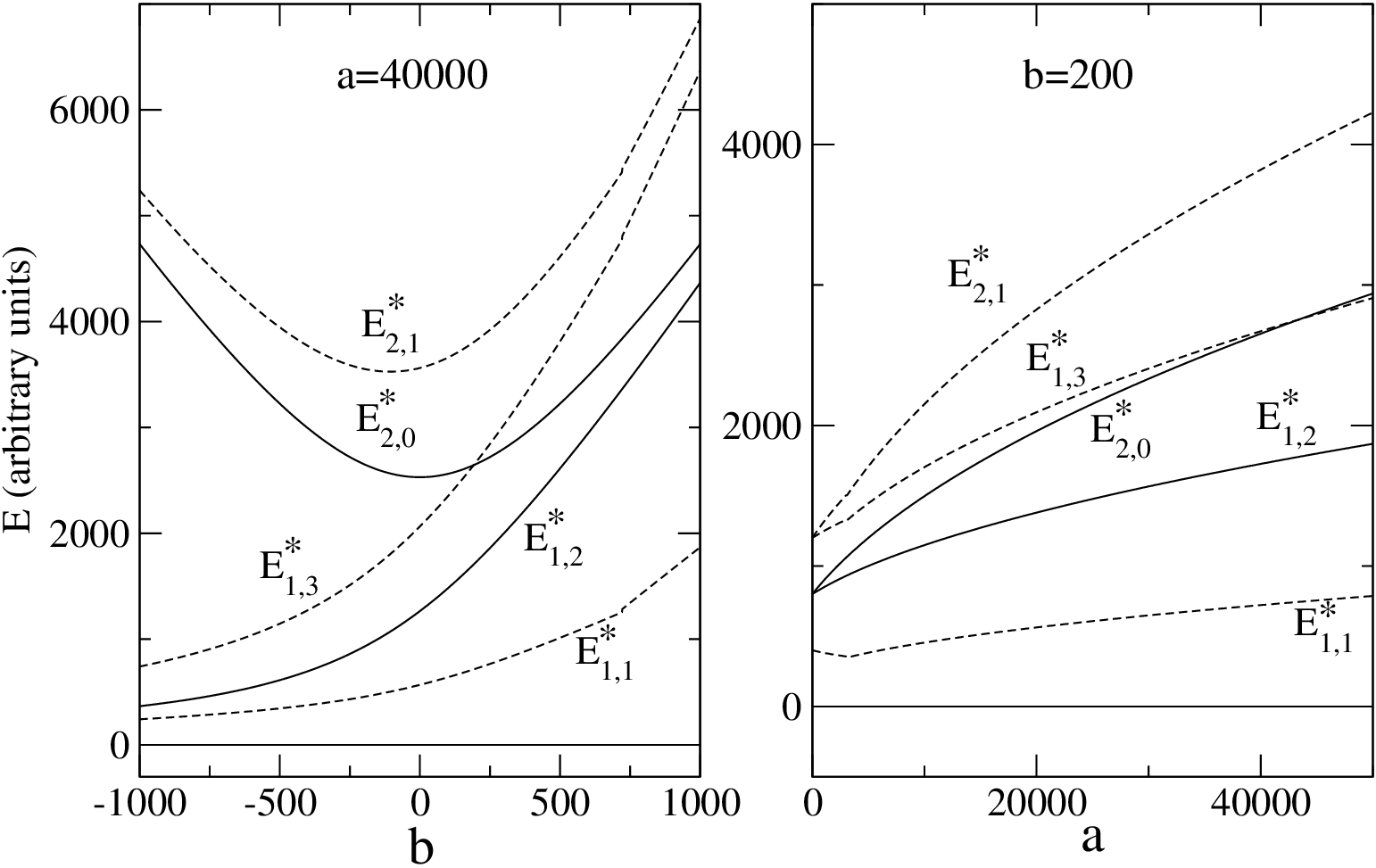}}}
\caption{Excitation energies $E^*_{\xi,\tau}=E_{\xi,\tau}-E_{1,0}$ 
  with $a=40000$ fixed  as a function of $b$ (left panel) and with
  $b=200$ fixed as a function of $a$ (right panel).}
\label{energies}
\end{figure}

Due to the relatively small difference in $c^+$ and 
$c^-$, the $\tau$-even and $\tau$-odd potentials have the 
same types of extrema at about the same $\beta$, except for some peculiar 
combinations of $a$ and $b$. Assuming that there are no complications 
of this kind, we can now return to the question of renormalizing the 
minima of the $\tau$-even and $\tau$-odd potentials. For 
$b>2(ac^{\pi})^{1/2}$, $\pi=+,\ -$ the minima of the two potentials 
will be $u_0^+$ and $u_0^-$ at $\beta=0$, so they coincide if 
$u_0^+=u_0^-$ holds. 
For $b<2(ac^{\pi})^{1/2}$ we can equate the minima of $u^+(\beta)$ and 
$u^-(\beta)$ if we set $u_0^+=0$ and 
\begin{equation}
u_0^-=(b^2-11a)(\beta^+_0)^2 -(b^2-13a)(\beta^-_0)^2 
+2ab[(\beta^+_0)^4 -(\beta^-_0)^4] 
+a^2[(\beta^+_0)^6 -(\beta^-_0)^6]\ ,
\label{u0}
\end{equation}
where the $\beta^{\pi}_0$ are obtained from (\ref{beta0}) with the 
choice of the ``+'' sign. With this the two potentials have their  
minima at the same energy, but they take on different values at 
the origin. 
Illustrations of the possible potential shapes are displayed 
in Fig. \ref{potentials}. 
Obviously, $u_0^-$ in (\ref{beta0}) also has to be 
added to the energies of the $\tau$-odd potential. 
Fig. \ref{energies} show the relative position of the energy 
levels $E_{\xi,\tau}$ as either $a$ or $b$ is varied and the other 
parameter is kept at a fixed value. 

The electric quadrupole transition rates can also be determined analytically 
by calculating the matrix elements of the transition operator 
\cite{wj56,fi00}
\begin{equation}
T^{({\rm E}2)}=t\alpha_{2\mu}
=t \beta\left[
D^{(2)}_{\mu ,0}\cos \gamma +2^{-1/2}(
D^{(2)}_{\mu ,2} + 
D^{(2)}_{\mu ,-2})\sin \gamma
\right]\ .
\label{te2}
\end{equation}
The radial integrals that appear in the $\beta$ variable in the matrix 
elements of $T^{({\rm E}2)}$ can again be determined using (\ref{sexintd}). 
In order to obtain the {\it total} matrix elements, one has to calculate 
also the components depending on $\gamma$ and  the Euler angles 
$\theta_i$. This can 
be done following the techniques described in Ref. \cite{bes}. These 
parts introduce certain selection rules not only for the angular 
momenta, but also for $\tau$. 

\section{Discussion}
\label{disc}

\begin{figure}
\resizebox{\columnwidth}{!}{\rotatebox{0}{\includegraphics{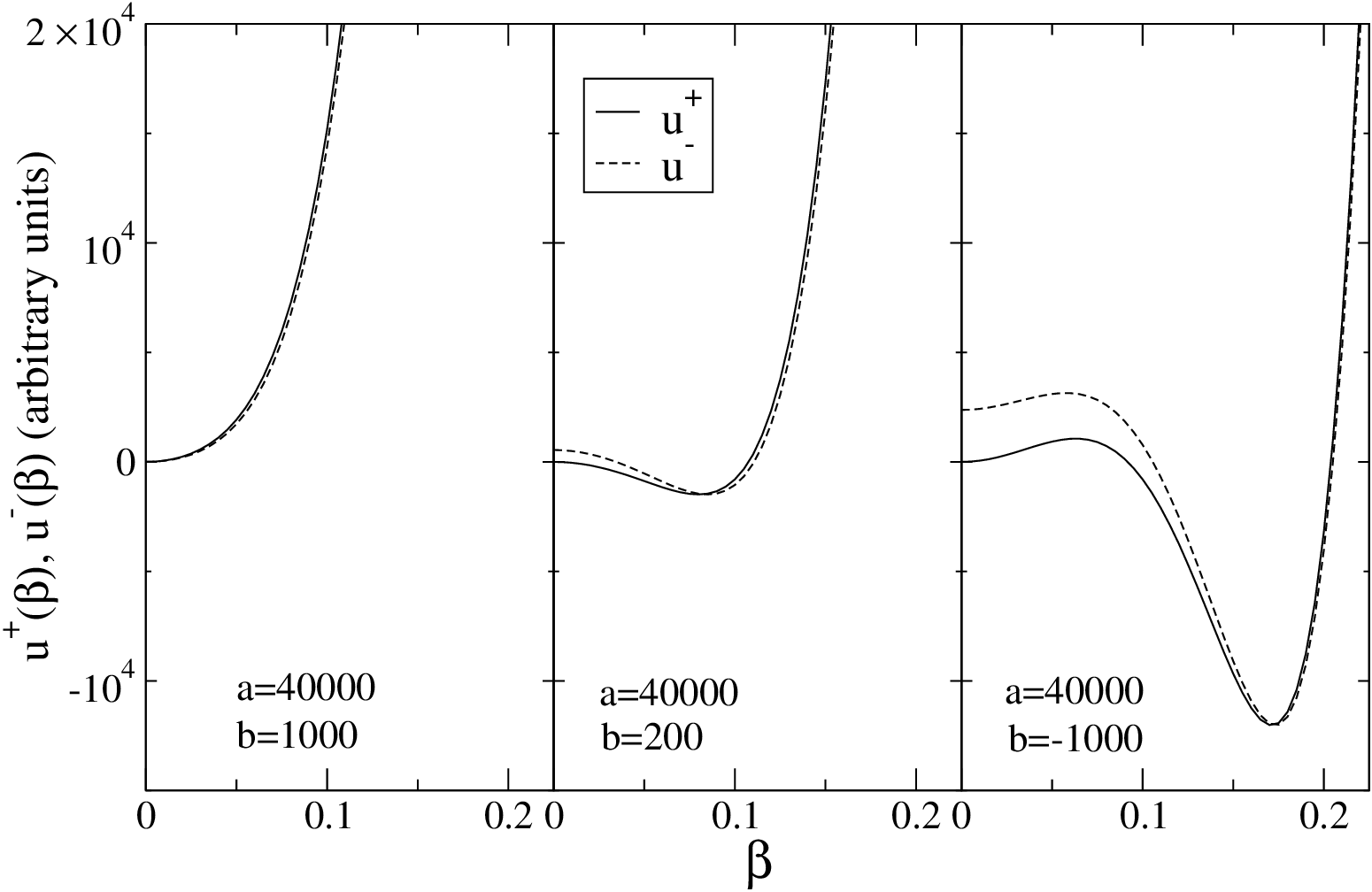}}}
\caption{Potentials $u^+(\beta)$ (full line) and $u^-(\beta)$ 
(broken line) for $a=40000$ and $b=1000$ (left panel), $b=200$ 
(middle panel) and $b=-1000$ (right panel). 
The lowest energy level appears in these potentials at 
$E_{1,0}=4633.57$, $73.35$ and $-9366.43$, respectively. 
}
\label{potentials}
\end{figure}

\begin{figure}
\resizebox{\columnwidth}{!}{\rotatebox{0}{\includegraphics{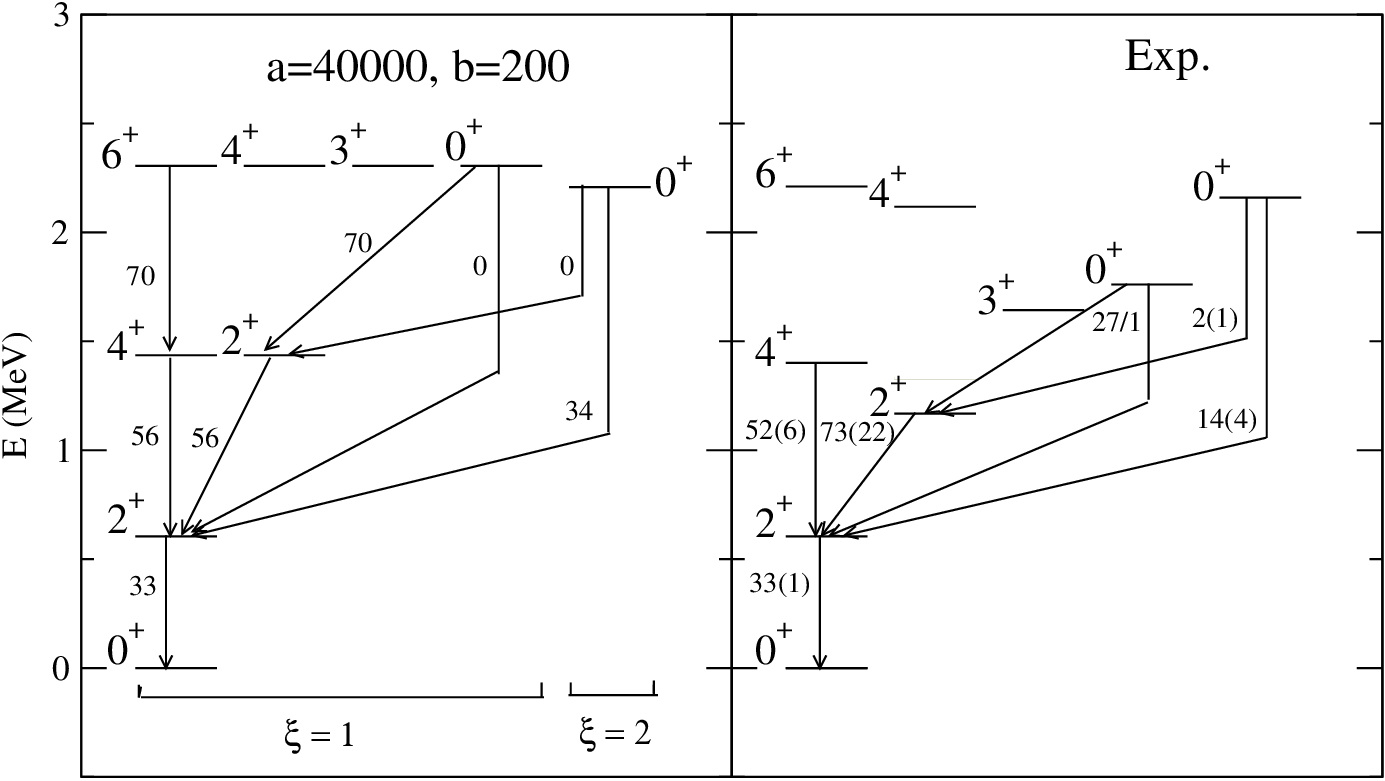}}}
\caption{The energy spectrum and the strength of some electric 
quadrupole transitions calculated with $a=40000$ and $b=200$ 
(left panel) and the corresponding data for $^{134}$Ba (right 
panel).} 
\label{134ba}
\end{figure}

In order to compare the main characteristics of the sextic oscillator 
as a $\gamma$-unstable potential with those of other potentials of 
this kind, we present calculations for a particular value of the 
parameters, $a=40000$ and $b=200$. These numbers were chosen such that 
the resulting energy spectrum approximates that of the $^{134}$Ba 
nucleus, the first candidate for E(5) symmetry \cite{casten}. We 
stress that our aim is not to reproduce the experimental data, rather 
to get a qualitative picture about the general performance of the 
model. The potentials $u^{\pm}(\beta)$ are displayed in the middle
panel of Fig. \ref{potentials}, while the energy eigenvalues are shown 
in Fig. \ref{134ba}, together with the corresponding experimental 
energy levels. Fig. \ref{134ba} also shows the calculated and the 
experimental $B({\rm E}2)$ values for transitions between the energy levels. 
Note that electric quadrupole transitions which change $\tau$ with 
more than one unit are zero if we use the transition operator 
(\ref{te2}), but finite $B({\rm E}2)$ strengths can be obtained if we apply 
terms of the next order (see e.g. \cite{frank}). 

In Table \ref{numbers} 
we summarize the ratio of the most important energy eigenvalues 
and those of the most characteristic $B({\rm E}2)$ transition rates obtained 
from the sextic oscillator with parameters $a=40000$, $b=200$, the 
infinite square well potential \cite{fi00} and the numerically 
solved $\beta^4$ potential \cite{beta4} together with the corresponding 
experimental values for $^{134}$Ba, whenever available. It is seen 
that the energy ratios corresponding to the E(5) symmetry systematically 
fall between the values of the $\beta^4$ potential and the sextic 
oscillator. The situation is less obvious for the ratio of the 
$B({\rm E}2)$ values: here the sextic oscillator and the infinite 
square well seem to yield similar ratios, while the numbers obtained 
from the $\beta^4$ potential are systematically higher. This might 
be due to the fact that the sextic oscillator potential goes to infinity 
steeper than the $\beta^4$ potential, so the asymptotic behaviour of its 
wavefunctions can be closer to that of the wavefunctions of the infinite
square well. 
Comparing the results with the experimental data for $^{134}$Ba we can
conclude that, at least in this case, the sextic oscillator allows a
better approximation than the other essentially parameter-free
potentials. We expect that this conclusion will be general due to the
flexible nature of the sextic potential whose shape is governed by two
parameters. In fact this potential can be used not just at the
critical point but it can be useful to model the full shape phase
transition from spherical to deformed $\gamma-$unstable nuclei by
changing the parameters $a$ and $b$.

Before closing, we mention some aspects of the sextic oscillator 
that might give further help in the analysis of nuclei near 
critical points. First we note that with $M=2$ in (\ref{sexpot}) the 
analysis can be extended to further states, such as $\phi_{1,4}$, 
$\phi_{1,5}$, $\phi_{2,2}$, $\phi_{2,3}$, $\phi_{3,0}$ and 
$\phi_{3,1}$. Second, the potential shape which contains both a 
local maximum and a minimum at $\beta>0$ might be useful in the 
description of nuclei with the so-called X(5) symmetry, which is 
thought to occur in the shape phase transition between the spherical 
and the axially deformed domain \cite{fi01}. Third, there are further 
quasi-exactly solvable potentials both with confining and non-confining 
nature \cite{qes}, which can also be considered in the Bohr Hamiltonian.

\begin{acknowledgments}
This work was supported by the OTKA grant No. T37502 (Hungary) and
by the Spanish MCyT under project No. BFM2002-03315 
\end{acknowledgments}

\newpage

\begin{table*}
\caption{\label{numbers} Ratios of some energy eigenvalues and 
electric quadrupole transition strengths from the sextic oscillator 
with $a=40000$, $b=200$, the infinite square well \cite{fi00} and the 
$\beta^4$ potential \cite{beta4}, together with the experimentally 
observed quantities for $^{134}$Ba. }
\begin{ruledtabular}
\begin{tabular}{lcccccc}
 & 
$\frac{E(4^+_{1,2})}{E(2^+_{1,1})}$ & 
$\frac{E(0^+_{2,0})}{E(2^+_{1,1})}$ & 
$\frac{E(6^+_{1,3})}{E(2^+_{1,1})}$ & 
$\frac{B(E2;4^+_{1,2}\rightarrow 2^+_{1,1})
 }{B(E2;2^+_{1,1}\rightarrow 0^+_{1,0})}$ & 
$\frac{B(E2;2^+_{2,0}\rightarrow 2^+_{1,1})
 }{B(E2;2^+_{1,1}\rightarrow 0^+_{1,0})}$ & 
$\frac{B(E2;0^+_{1,3}\rightarrow 2^+_{1,2})
 }{B(E2;2^+_{1,1}\rightarrow 0^+_{1,0})}$ \\
\hline
sextic osc. & 2.39 & 3.68 & 3.70 & 1.70 & 1.03 & 2.12 \\
E(5) & 2.20 & 3.03 & 3.59 & 1.68 & 0.86 & 2.21 \\
$\beta^4$ & 2.09 & 2.39 & 3.27 & 1.82 & 1.41 & 2.52 \\
$^{134}$Ba (exp.) & 2.31 & 3.57 & 3.65 & 1.56(18) & 0.42(12) &  \\
\end{tabular}
\end{ruledtabular}
\end{table*}

\newpage

\begin{thebibliography}{99}

\bibitem{bm} A. Bohr and B. Mottelson, {\it Nuclear Structure} 
             (Benjamin, Reading, MA, 1975) Vol. II.

\bibitem{ibm} F. Iachello and A. Arima, {\it The Interacting Boson Mode} 
             (Cambridge University Press, Cambridge, 1987).

\bibitem{Die} A.E.L. Dieperink, O. Scholten and F. Iachello, 
             {\it Phys. Rev. Lett.} {\bf 44}, 1747 (1980). 

\bibitem{Feng} D.H. Feng, R. Gilmore and S.R. Deans, 
             {\it Phys. Rev. C} {\bf 23}, 1254 (1981). 

\bibitem{Cas}  E. L\'opez-Moreno and O. Casta\~{n}os, 
             {\it Phys. Rev. C} {\bf 54}, 2374 (1996). 

\bibitem{Jo1}  J. Jolie et. al., 
             {\it Phys. Rev. Lett.} {\bf 89}, 182502 (2002).

\bibitem{Ar3}  J.M. Arias, J. Dukelsky and J.E. Garc\'{\i}a-Ramos, 
             {\it Phys. Rev.  Lett.} {\bf 91}, 162502 (2003).

\bibitem{fi00} F. Iachello, 
              {\it Phys. Rev. Lett.} {\bf 85}, 3580 (2000).
 
\bibitem{fi01} F. Iachello, 
              {\it Phys. Rev. Lett.} {\bf 87}, 052502 (2001).

\bibitem{fi03} F. Iachello, 
              {\it Phys. Rev. Lett.} {\bf 91}, 132502 (2003).

\bibitem{wj56} L. Wilets and M. Jean, 
              {\it Phys. Rev.} {\bf 102}, 788 (1956).

\bibitem{elliott} J. P. Elliott, J. A. Evans and P. Park, 
                  {\it Phys. Lett.} {\bf 169B}, 309 (1986).

\bibitem{fortunato} L. Fortunato and A. Vitturi, 
                  {\it J. Phys. G} {\bf 29}, 1341 (2003). 

\bibitem{as70} M. Abramowitz and I. A. Stegun, {\it Handbook of 
        Mathematical Functions} (Dover, New York, 1970). 

\bibitem{qes} A. G. Ushveridze, {\it Quasi-exactly solvable models in 
       quantum mechanics} (IOP Publishing, Bristol, 1994). 

\bibitem{bes} D. R. B\`es, {\it Nucl. Phys.} {\bf 10}, 379 (1959). 

\bibitem{casten} R. F. Casten and N. V. Zamfir,
              {\it Phys. Rev. Lett.} {\bf 85}, 3584 (2000).
 
\bibitem{frank} A. Frank, C. E. Alonso and J. M. Arias,
              {\it Phys. Rev. C} {\bf 65}, 014301 (2001).
 
\bibitem{beta4} J. M. Arias, C. E. Alonso, A. Vitturi, J. E. Garcia-Ramos, 
                J. Dukelsky and A. Frank, 
                {\it Phys. Rev. C} {\bf 68}, 041302(R) (2003).

\end{thebibliography}
\end{document}